\documentclass[twocolumn,showpacs]{revtex4}

\usepackage{amssymb}
\usepackage{graphicx}
\usepackage{amsmath}
\usepackage[dvips]{epsfig}

\makeatletter
\def\btt#1{\texttt{\@backslashchar#1}}
\DeclareRobustCommand\bblash{\btt{\@backslashchar}}
\makeatother

\begin{document}

\preprint{Manuscript}
\title{Thermal conductivity of epitaxial graphene nanoribbons on SiC: effect of substrate }
\author{Zhi-Xin Guo$^{1}$$^{\ast }$, J. W. Ding$^{1}$ and Xin-Gao Gong$^{2}$}
\affiliation{$^{1}$Department of Physics, Xiangtan University,
Xiangtan, Hunan 411105, China\\  $^{2}$ Surface Physics Laboratory
and Department of Physics, Fudan University, Shanghai 200433, China}
\email{zxguo08@gmail.com}
\date{\today }

\begin{abstract}
We study the effect of SiC substrate on thermal conductivity of epitaxial
 graphene nanoribbons (GNRs) using the
nonequilibrium molecular dynamics method. We show that the substrate has strong interaction
with single-layer GNRs during the thermal transport, which largely reduces the thermal conductivity.
The thermal conductivity characteristics of suspended GNRs are well preserved in the second GNR layers of bilayer GNR,
which has a weak van der Waals interaction with the underlying structures.
The out-of-plane phonon mode is found to play a critical role on
the thermal conductivity variation of the second GNR layer induced by the underlying structures.
\end{abstract}
\maketitle

Graphene and Graphene nanoribbons (GNRs) are thought to be ideal materials for
 nanoelectronics due to their outstanding electronic and thermal
 properties\cite{1,2,3,4,5,6}. In the production, the graphene nanomaterials can
 be either prepared by mechanical
exfoliation from graphite\cite{7,8} or by epitaxial growth on SiC substrate\cite{9,10}.
However, the mechanical method is quite delicate and time consuming,
which makes it unapplicable in the industry. The epitaxial growth
method is nowadays commonly accepted to represent a viable method of
controllable growth for the fabrication of high quality graphene
wafers\cite{11}. The electronic properties of epitaxial graphene on SiC substrate had
been extensively studied\cite{10,12,13,14}. It was found the electronic properties of
graphene can be well preserved in both the single-layer (SL) graphene on
SiC (000$\bar{1}$) (C-terminated)\cite{10}, and the second layer of bilayer (BL) graphene on
 SiC(0001)(Si-terminated)\cite{12,13}, having
great application potential of epitaxial graphene in the nanoelectronics. Since the heat
removal is a crucial issue in the nanoelectronic industry, the
thermal conduction property of epitaxial graphene and GNRs becomes
particularly important to its application in the nanoelectronics.

During the last two years, the thermal conductivity of exfoliated
graphene on different substrates (SiO2, Cu) has been extensively
studied, where only weekly coupled graphene-substrate interaction exists\cite{15,16,17,18,19}.
Different from the exfoliated graphene case, the
epitaxial graphene-substrate interaction is much more complicated,
and the geometry can even be distorted by the substrate \cite{12,20}.
Thus the thermal conductivity of epitaxial graphene and GNRs is expected to be very different from the
exfoliated ones.

In this work, we use the nonequilibrium molecular dynamics (NEMD)\cite{6,21,22,23}
method to study the thermal conductivity of epitaxial GNRs on
4H-SiC (000$\bar{1}$) and (0001) surface.
On the (000$\bar{1}$) surface, both covalently bonded and weakly coupled
GNR-substrate interaction conditions that were observed by the
experiments are considered. On the (0001) surface, we consider both the SL and BL GNR-substrate
interaction cases. In the BL GNR, we concentrate on the
thermal conductivity of the second layer since the first
layer is expected to have similar thermal conductivity as the SL GNR.
Two typical GNRs, i.e., armchair GNR (AGNR) and
zigzag GNR (ZGNR), are considered, and we refer to AGNR/ZGNR with N dimer lines in width as N-AGNR/N-ZGNR
 for convenient representation\cite{24}.

The 4H-SiC substrate is modeled with four alternating Si and
C atomic layers. One Si-C layer at the bottom of the
sample is fixed. GNRs are placed on top of
the SiC substrate, with infinite length alone the X direction and
finite width along Y direction. The smallest cell of the GNR-SiC
system is of 10.12 nm in length (X direction) and 2.15 nm (around 1
nm) for the SiC (GNR) in width (Y direction), containing 2128 SiC
atoms and 320 (368) 4-ZGNR (8-AGNR) atoms, respectively.

In the geometry optimization, periodic boundary conditions are
applied both along the X and Y directions. We use the
Tersoff\cite{25} potential to describe the C-C and C-Si bonded
interactions, and the non-bonded van der Waals interaction is
described by the Lennard-Jones (LJ) potential\cite{26}, which is only nonzero
after the Tersoff covalent potential goes to zero. The coupling
between the long-range LJ potential and the short-range Tersoff
potential is described by a cubic spline function\cite{27}.

In the NEMD simulation, we employ the velocity Verlet method to
integrate equations of motion with a fixed time step of 1 fs. Fixed
boundary condition is applied along the X direction, where the
outmost layers of each end of GNR (SiC) are fixed. Next to the
boundaries, the adjacent 1 nm-long GNR (SiC) layers are coupled to the
Nos\'{e}-Hoover\cite{28} thermostats with temperatures 310 and 290 K,
respectively. The thermal conductivity of GNRs $\kappa$ is then
calculated from the Fourier law,
\begin{equation}
\kappa=-\frac{J}{\nabla T\cdot S}\text{ ,}  \tag{1}
\end{equation}
where $J$ is the heat flux from the thermostats to the system, which
is obtained from the Green-Kubo relation\cite{29,30}. $\nabla T$ is the
temperature gradient in the length direction, which is defined as $\nabla
T=(T_{L}-T_{R})/L$, where $T_{L}$, $T_{R}$ are the temperature of
thermostats at the two ends, and $L$ is GNR length. $S$ is the
cross-section area. In this work, we choose the interplanar spacing of
graphite 3.35 {\AA} as the GNRs' thickness. Moreover, all results given in
this paper are obtained by averaging about 5 ns after a sufficient
long transient time (5 ns) when a nonequilibrium stationary state
is set up.

\begin{figure}[tbp]
\includegraphics[scale=0.3,angle=90]{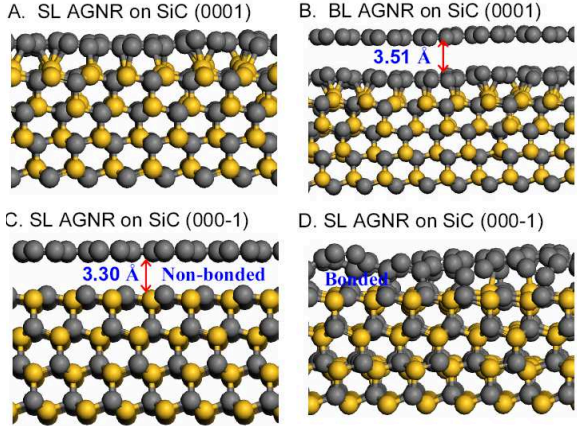}
\caption{ Optimized structures of 8-AGNR on SiC substrate for 4 cases:
 A, single-layer GNR on SiC (0001); B, bilayer GNR on SiC
(0001); C, single-layer GNR on SiC (000$\bar{1}$) with weakly coupled interaction; D, single-layer GNR on SiC
(000$\bar{1}$) with covalently bonded interaction. Carbon atoms are represented
by gray balls; Si, by yellow balls.}
\end{figure}

The following
4 kinds of epitaxial GNRs on SiC are considered: A, SL GNR on SiC (0001);
B, BL GNR on SiC(0001); C, SL GNR on SiC (000$\bar{1}$) with the
weakly coupled interaction; D, SL GNR on SiC
(000$\bar{1}$) with the covalently bonded interaction.
Before geometry optimization, the initial distance between GNR and SiC
surface was set to be 2.3 {\AA} in cases A and B, 3.0 {\AA} in case C,
and 2.0 {\AA} in case D. In Fig. 1, we show the optimized structures of
8-AGNR on SiC for the above 4 cases. As one can see, the SL GNRs can either
 covalently bonded or weakly coupled to the substrate. In the covalently bonded case (case A, D), the formed
C-Si (C-C) covalent bond between GNR and substrate has a length
around 2.15 (1.60) {\AA} in the middle region, being consistent with the bond
length of the graphene-SiC system in previous reports \cite{12,13,31}. While, in the
edge region the C-Si (C-C) bond length is only about 2.10 (1.50) {\AA}, shorter
than the bond length in the middle region, indicating that the edge atoms make the interaction between
GNR and substrate stronger than that between graphene and substrate.
In the weakly coupled case (case C), the mean distance between GNR and SiC surface is 3.30 {\AA}
, similar with the graphene case\cite{32}.

In the BL GNR (case B), while, the mean distance between
the first and the second layer is 3.51 {\AA},
obviously larger than the interlayer  distance of the suspended BL graphene.
 The larger interlayer distance comes from the larger shear modules
 of GNRs\cite{33} and the large ripples formed on first GNR layer due to
its strong interaction with SiC surface\cite{34}.
 Compared with the BL graphene, the interlayer force in BL GNR
is smaller due to its finite width, while its shear modules is
much larger. Thus, unlike the BL graphene,
 the weak interlayer van der Walls
 force can't make the second GNR layer follows the large ripples of the first
 GNR layer, and the second GNR layer still keeps a relatively planar structure. The large
 interlayer distance appears at the trough of the ripple in first GNR layer.

\begin{figure}[tbp]
\includegraphics[scale=0.3,angle=-90]{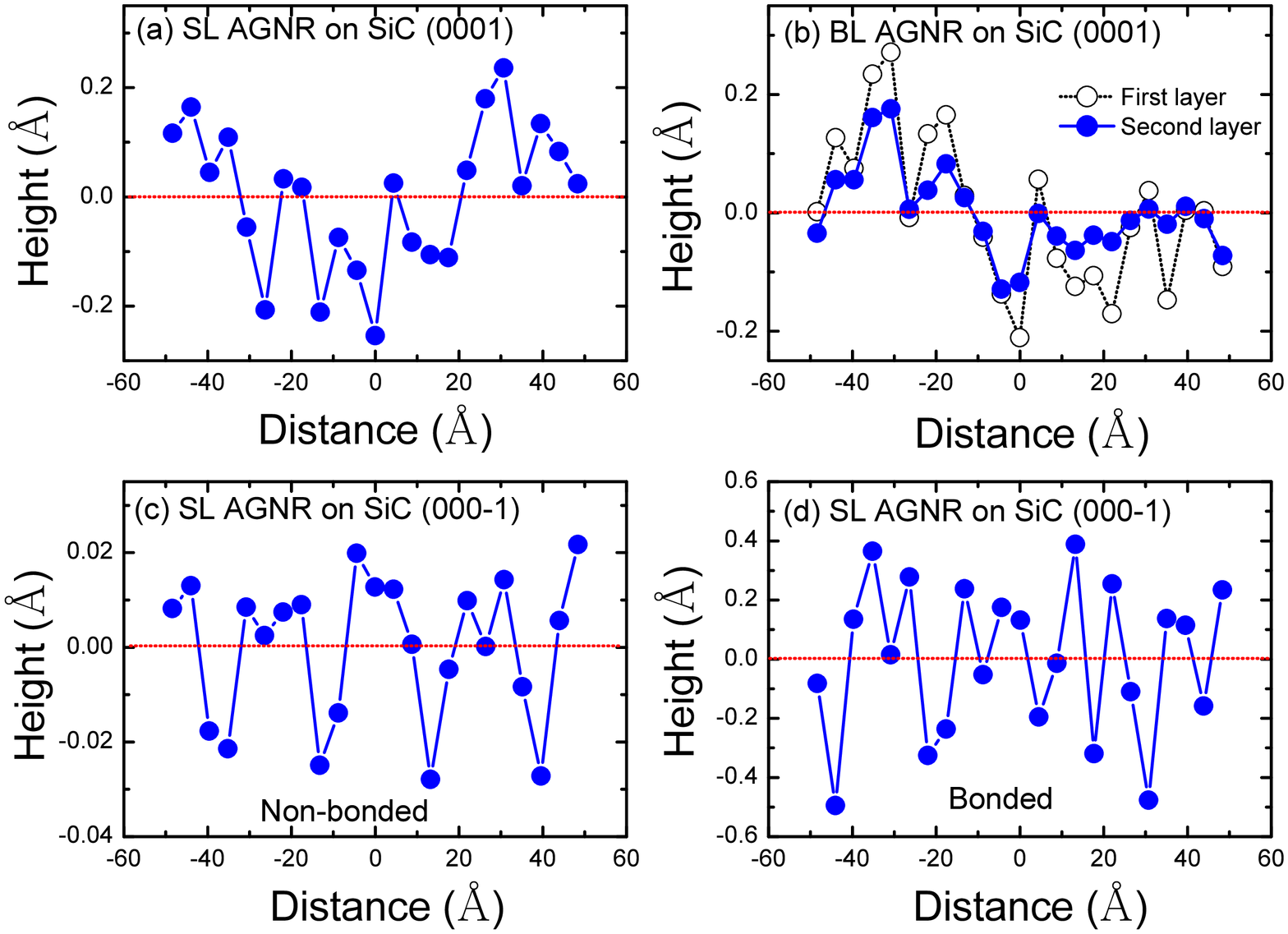}
\caption{ The height profile of GNRs on SiC along the length direction for
cases A (a), B (b), C (c), and D (d).}
\end{figure}

The corresponding height profile of GNRs on SiC along the length direction is shown in Fig. 2. From the figure,
 the ripples of GNRs that covalently bonded with SiC are
obviously larger than that of GNR weakly coupled with SiC. The
smallest and largest ripples appear in case C and D, respectively, both of which are
on SiC (000$\bar{1}$).
Moreover, in the BL GNR the
 ripple of the second GNR layer is obviously smaller than that of first GNR
 layer (Fig. 2(b)), which corresponds to a larger interlayer distance as discussed above.

\begin{figure}[tbp]
\includegraphics[scale=0.3,angle=90]{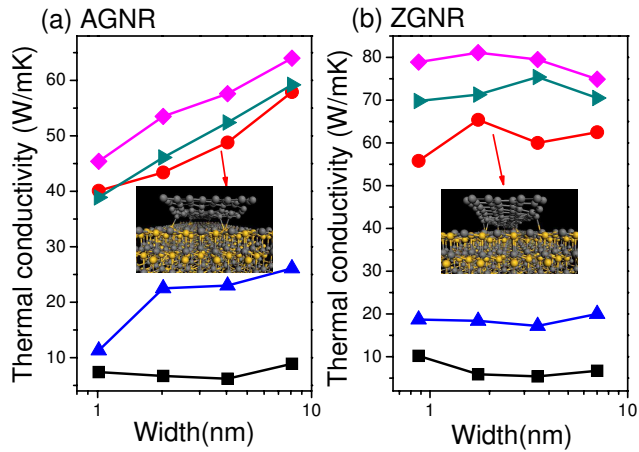}
\caption{Width dependence of thermal conductivity of both
AGNRs (a) and ZGNRs (b) on SiC in cases A (blue uptriangle), B (cyan righttriangle ),
C (red circle), and D (black square), with comparison of the suspended GNRs (magenta diamond). }
\end{figure}

Fig. 3 shows the width dependence of thermal conductivity of both
AGNRs and ZGNRs on SiC. As shown in the figure, the thermal conductivity of GNRs covalently bonded with SiC is
very different from that weakly coupled with SiC. In the covalently bonded case, thermal conductivity of GNRs is very low (below 30 W/mK),
and has little variation when the width gets larger than 2 nm. This shows that the strong covalent Si-C/C-C bonds formed between GNR and
 SiC surface have destroyed the intrinsic thermal conductivity of GNRs. In addition, the thermal conductivity of GNR on SiC (000$\bar{1}$)
 (case D) is distinctly smaller than that on the (0001) surface (case A). This is attributed to the larger GNR ripples
 formed on the (000$\bar{1}$) surface than that on (0001) surface (Fig. 2), which would induce stronger phonon scattering.
Different from the covalently bonded cases, thermal conductivity of GNRs weakly coupled with SiC is much higher (case C).
However, we find that some C-Si covalent bonds are newly formed
between the edge of SL GNR and SiC(000$\bar{1}$)
surface during the NEMD simulation (inset of Fig. 3),
which is attributed to the large atomic amplitude of GNR in the direction
perpendicular to the SiC surface when the system is thermostated to 300 K.
The covalent bonds would induce additional edge localized phonon scattering on the GNR, and thus reduce the thermal conductivity.
Since the number of newly formed covalent bonds is proportional to the length of GNR, compared with that of the suspended GNR,
 we expect the thermal conductivity will be largely reduced when the GNR length gets to $\mu m$-scale (experimental length).

From Fig. 3, it is definite that, the thermal conductivity of the second GNR layer in a BL GNR (case B)
 is most close to that of the suspended GNR.
 Also, the width-dependence of thermal conductivity is very similar to each other:
 With the width increasing, the thermal conductivity of AGNR monotonously increases,
 while the thermal conductivity of ZGNR increases first and then decreases\cite{6}. This implies the weak van der Waals interaction
from underlying structures (both the first GNR layer and SiC substrate) does not break the intrinsic thermal conductivity of the second GNR layer.

\begin{figure}[tbp]
\includegraphics[scale=0.3,angle=-90]{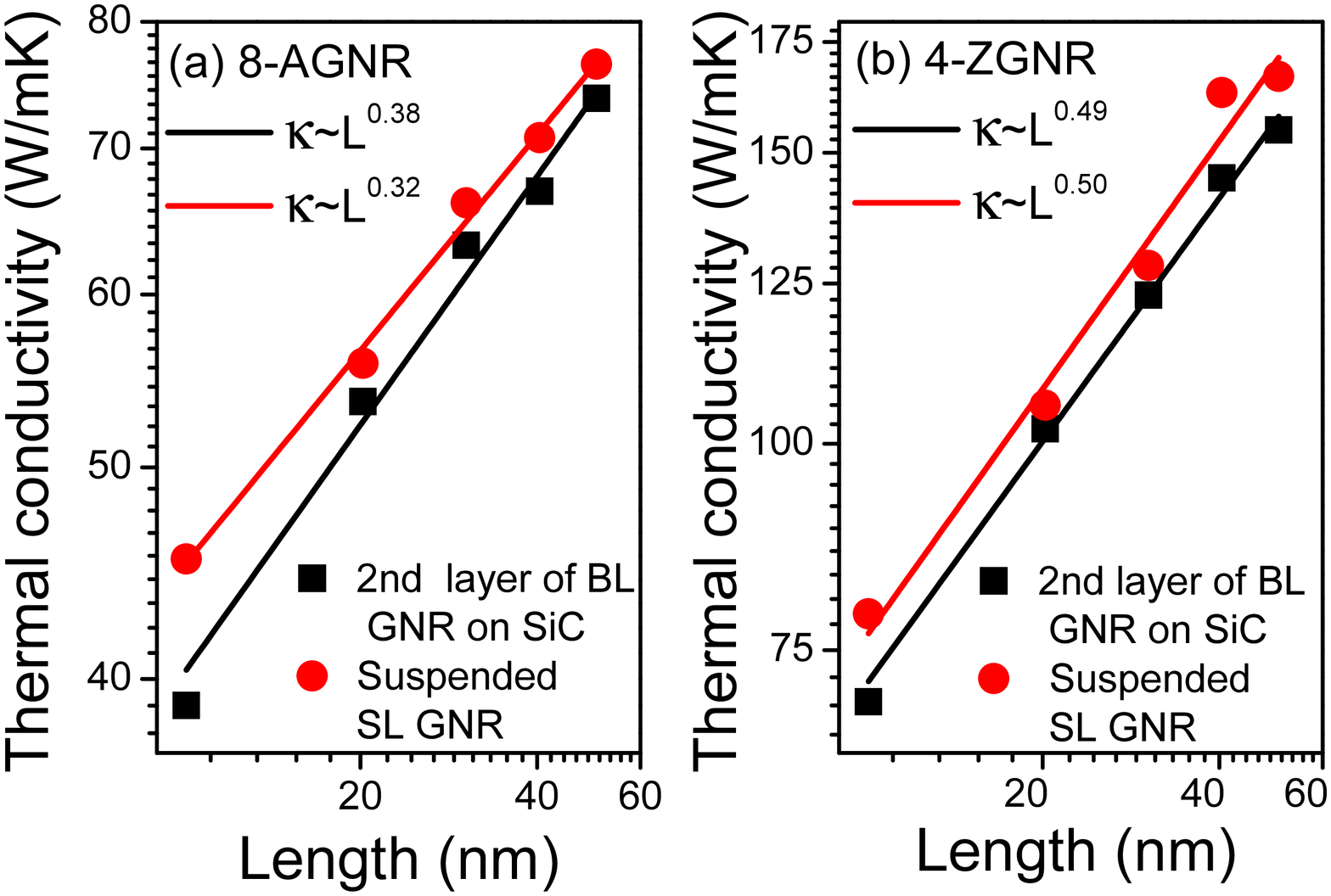}
\caption{Thermal conductivity $\kappa$ vs the length $L$ in log-log scale for the second 8-AGNR (a), 4-ZGNR (b) layers in BL GNR,
 with comparison of the suspended GNRs.
In both cases, $\kappa \sim L^{\beta}$, with $\beta$ being close to that of the suspended GNRs.}
\end{figure}

 In Fig. 4
we show the length dependence of thermal conductivity of the second GNR layers of BL GNR with comparison of the suspended GNRs. Similar with that of
the suspended GNRs and carbon nanotubes\cite{6,35},
the thermal conductivity of the second 8-AGNR (4-ZGNR) layer monotonously increases with the length increasing and follows
 a power law of $\kappa \sim L^{\beta}$, with $\beta$=0.38 (0.49). The value of $\beta$ is very close to that of the suspended GNR
 ($\beta$=0.32 (0.50)), showing that the ballistic and 1D thermal transport characters\cite{35,36,37}
  are well preserved in the second GNR layer.

\begin{table}
 \caption{\label{tabone} Thermal conductivity $\kappa$ of 8-ZGNRs with/without the out-of-plane
 vibrational constraint. The GNR length is kept at 10 nm.}
\begin{tabular}{cccccccc}
  \hline
  GNR type & ~~~~Constraint & ~~~~$\kappa$~(W/mK)~~~~  \\
    \hline
 Suspended SL GNR &~~~~ Free & ~~~81.1  \\
 Suspended SL GNR & ~~~~Constraint & ~~~77.7 \\
2nd layer of BL GNR & ~~~~ Free & ~~~71.3 \\
2nd layer of BL GNR & ~~~~ Constraint & ~~~79.1 \\
  \hline
\end{tabular}
\end{table}
To further explore the substrate effect on thermal conductivity of the second GNR layer of BL GNR on SiC,
we freeze the out-of-plane atomic vibration of the second GNR layer and recalculate the thermal conductivity with comparison
of the suspended SL GNR.
For simplicity, here we only present the calculated thermal conductivity
of ZGNRs, similar result is also obtained for AGNRs. As shown in Table 1, freezing the out-of-plane atomic vibration
 decreases the thermal conductivity of the suspended SL GNR, while it considerably increases
the thermal conductivity of the second GNR layer of BL GNR.

This illustrates two facts.
One is that the out-of-plane phonon mode in suspended SL GNR has a positive contribution to the thermal
transport, although it is not dominating. The other is that the thermal conductivity reduction of the second GNR
layer of BL GNR mainly comes
from the coupling between its out-of-plane phonon mode and the phonon modes of the underlying structures (including phonon modes of both
first GNR layer\cite{38} and SiC surface).
It should be mentioned that, the thermal conductivity of the second GNR layer (79.1 W/mK) in BL GNR is even higher than
that of the suspended SL GNR (77.7 W/mK) when the out-of-plane atomic vibration is frozen. This interesting phenomenon
indicates an increment of thermal conductivity can be realized
 through the substrate coupling without the out-of-plane phonon.
 Recently, through a coupled atomic chain model, we have clarified that there exists a competitive mechanism on thermal conductivity
in a coupling system: the phonon resonance effect that
decreases thermal conductivity and phonon-band-up-shift effect
that increases thermal conductivity\cite{23}. In this case, the phonon resonance effect mainly comes from the coupling
between out-of-plane phonon mode of the second GNR layer and the phonon modes of the first GNR layer and SiC substrate.
When the out-of-plane atomic vibration is frozen, the phonon resonance effect would be largely reduced, and the phonon-band-up-shift effect
become dominated. Thus the thermal conductivity can be increased by the coupling. The results further confirm the
existence of two competitive effects between the thermal conductive material and substrate.

In summary, we have investigated the thermal conductivity of epitaxial SL and BL
GNRs on SiC substrate using the NEMD method. For the SL GNR, both covalently bonded and weakly coupled
GNR-substrate interaction conditions that were observed by experiments are considered. The thermal conductivity of SL GNRs
in the covalently bonded condition is particularly low, while it is much higher in the weakly coupled condition.
 However, there appear some C-Si covalent
 bonds between the edge of SL GNR and SiC surface in the weakly coupled condition during the thermal transport,
 which induces additional edge localized phonon scattering and is expected to largely reduce the thermal conductivity
 when the GNR length gets to $\mu m$-scale.
The second GNR layer of BL
 GNRs is found to have the highest thermal conductivity among all the epitaxial GNRs, and
keeps much the same thermal conductivity characteristics as the suspended GNR.
We find that the out-of-plane phonon mode of the
 second GNR layer plays a critical role on the thermal conductivity variation induced by the underlying structures. The
existence of two competitive effects between thermal conductive material and substrate is further confirmed.
We expect the present study can be helpful to the forthcoming applications of epitaxial graphene nanomaterials
in the nanoelectronics.

This work was supported by the Start-up funds (No. 10QDZ11), Scientific Research Fund (10XZX05) of Xiangtan
University, and PCSIRT (IRT1080).



\begin{references}

\bibitem{1} A. K. Geim, Science 324, 1530 (2009).

\bibitem{2} A. H. Castro Neto, F. Guinea, N. M. R. Peres, K. S. Novoselov, and A. K. Geim, Rev. Mod. Phys. 81, 109 (2009).

\bibitem{3} A. A. Balandin, S. Ghosh, W. Bao, I. Calizo, D. Teweldebrhan, F. Miao,and C. N. Lau, Nano Lett. 8, 902 (2008).

\bibitem{4} A. A. Balandin, Nature Mater. 10, 569 (2011).

\bibitem{5} J. Hu, X. Ruan, and Y. P. Chen, Nano Lett. 9, 2730 (2009).

\bibitem{6} Z. X. Guo, D. Zhang, and X. G. Gong, Appl. Phys. Lett. 95, 163103 (2009).

\bibitem{7} K. S. Novoselov, A. K. Geim, S. V. Morozov, D. Jiang, M. I. Katsnelson, I. V. Grigorieva, S. V. Dubonos, and A. A. Firsov,
            Nature 438, 197 (2005).

\bibitem{8} Y. Zhang, Y. W. Tan, H. L. Stormer, and P. Kim, Nature 438, 201 (2005).

\bibitem{9} W. A. de Heer, C. Berger, X. S. Wu, P. N. First, E. H. Conrad, X. B. Li, T. B. Li, M. Sprinkle, J. Hass, M. L. Sadowski, M. Potemski,
            and G. Martinez, Solid State Commun. 143, 92 (2007).

\bibitem{10} K. V. Emtsev, F. Speck, T. Seyller, L. Ley, and J. D. Riley, Phys. Rev. B 77, 155303 (2008).

\bibitem{11} C. Dimitrakopoulos, Y. M. Lin, A. Grill, D. B. Farmer, M. Freitag, Y. Sun, S. J. Han,
             Z. Chen, K. A. Jenkins, Y. Zhu, Z. Liu, T. J. McArdle, J. A. Ott, R. Wisnieff,
             and P. Avouris, J. Vac. Sci. Technol. B 28, 985 (2010).

\bibitem{12} A. Mattausch,  and O. Pankratov, Phys. Rev. Lett. 99, 076802 (2007).

\bibitem{13} F. Varchon, R. Feng, J. Hass, X. Li, B. NgocNguyen, C. Naud, P. Mallet,
             J. Y. Veuillen, C. Berger, E. H. Conrad, and L. Magaud, Phys. Rev. Lett. 99, 126805 (2007).

\bibitem{14} S. Kim, J. Ihm, H. J. Choi, and Y. W. Son, Phys. Rev. Lett. 100, 176802 (2008).

\bibitem{15} J. H. Seol, I. Jo, A. L. Moore, L. Lindsay, Z. H. Aitken, M. T. Pettes,
X. Li, Z. Yao, R. Huang, D. A. Broido, N. Mingo, R. S. Ruoff, and
L. Shi, Science 328, 213 (2010).

\bibitem{16} W. Jang, Z. Chen, W. Bao, C. N. Lau, and C. Dames, Nano Lett. 10, 3909 (2010).

\bibitem{17} Y. K. Koh, M. H. Bae, D. G. Cahill, and E. Pop, Nano Lett. 10, 4363 (2010).

\bibitem{18} Z. Wang, R. Xie, C. T. Bui, D. Liu, X. Ni, B. Li, and T. L. J. Thong, Nano Lett. 11, 113 (2011).

\bibitem{19} Z. Y. Ong, and E. Pop, Phys. Rev. B 84, 075471 (2011).

\bibitem{20} V. Sorkin, and Y. W. Zhang,  Phys. Rev. B 81, 085435 (2010).

\bibitem{21} Z. X. Guo, D. Zhang, Y. T. Zhai, and X. G. Gong, Nanotechnology 21, 285706 (2010).

\bibitem{22} Z. X. Guo, and X. G. Gong, Front. Phys. 4, 389 (2009).

\bibitem{23} Z. X. Guo, D. Zhang, and X. G. Gong, Phys. Rev. B 84, 075470 (2011).

\bibitem{24} Y. W. Son, M. L. Cohen, and S. G. Louie, Phys. Rev. Lett. 97, 216803 (2006).

\bibitem{25} J. Tersoff, Phys. Rev. B 39, 5566 (1989).

\bibitem{26} L. A. Girifalco, M. Hodak, and R. S. Lee, Phys. Rev. B 62, 13104 (2000).

\bibitem{27} Z. Mao, A. Garg, and S. B. Sinnott, Nanotechnology 10, 273 (1999).

\bibitem{28} S. Nos\'{e}, J. Chem. Phys. 81, 511 (1984); W. G. Hoover, Phys. Rev. A 31, 1695 (1985).

\bibitem{29} P. K. Schelling, S. R. Phillpot, and P. Keblinski Phys. Rev. B 65, 144306 (2002).

\bibitem{30} N. Yang, G. Zhang, and B. Li, Nano Lett. 8, 276 (2008).

\bibitem{31} Y. Qi, S. H. Rhim, G. F. Sun, M. Weinert, and L. Li, Phys. Rev. Lett. 105, 085502 (2010).

\bibitem{32} F. Hiebel, P. Mallet, J. Y. Veuillen, and L. Magaud, Phys. Rev. B 83, 075438 (2011).

\bibitem{33} R. Faccio, P. A. Denis, H. Pardo, C. Goyenola,
and \'{A}. W. Mombr\'{u}, J. Phys.: Condens. Matter 21, 285304 (2009).

\bibitem{34} F. Varchon, P. Mallet, J. Y. Veuillen, and L. Magaud, Phys. Rev. B 77, 235412 (2008).

\bibitem{35} S. Maruyama, Physica B 323, 193 (2002).

\bibitem{36} E. Enrique, J. Lu, and B. I. Yakobson, Nano Lett. 10, 1652 (2010).

\bibitem{37} G. Zhang, and B. Li, J. Chem. Phys. 123, 114714 (2005).

\bibitem{38} H. Y. Cao, Z. X. Guo, H. J. Xiang, and X. G. Gong, Phys. Lett. A 376, 525 (2012).

\end{references}
\end{document}